\documentclass[aps,prl,reprint,superscriptaddress,amsmath,amssymb]{revtex4-2}

\usepackage{graphicx}
\usepackage{bm} 
\usepackage{braket} 
\usepackage[colorlinks=true]{hyperref}
\hypersetup{linkcolor=blue,citecolor=blue,urlcolor=blue}

\begin{document}

\title{Exact Quench Dynamics from Thermal Pure Quantum States}

\author{Hui-Huang Chen}
\email{chenhh@jxnu.edu.cn}
\affiliation{School of Physics, Jiangxi Normal University, Nanchang 330022, China}
\affiliation{SISSA and INFN Sezione di Trieste, via Bonomea 265, 34136 Trieste, Italy}

\date{\today}

\begin{abstract}
We present an exact solution for entanglement entropy for the real-time dynamics following a quench from a thermal pure quantum (TPQ) state in a free-fermion system. In contrast to the usual linear growth and saturation behavior, the entanglement entropy exhibits a characteristic double-plateau structure. We establish this behavior through three complementary approaches: an exact conformal field theory calculation on the Klein bottle, finite-size Gaussian-state simulations, and a quasiparticle picture that becomes quantitatively accurate in the scaling regime.
\end{abstract}

\maketitle

{\em Introduction}. Understanding the non-equilibrium dynamics of isolated quantum systems is a central challenge in statistical mechanics and quantum gravity \cite{Srednicki1994,rigol2008,polkovnikov2011,sekino2008,liu2014}. Usually, analytical and numerical studies rely on low-entanglement initial states (e.g., ground states or short-range product states), from which the celebrated linear growth and subsequent saturation of entanglement emerge \cite{calabrese2005, alba2017, nahum2017}. In contrast, thermal pure quantum (TPQ) states are usually constructed from random states and already exhibit volume-law entanglement and are locally indistinguishable from Gibbs ensembles \cite{sugiura2012}. The entanglement dynamics from such initial states often involves exponential computational complexity, making exactly solvable models for this process very rare. Distinct from the traditional random-state-based TPQ states used for numerical calculations, recent literature has introduced highly structured TPQ states \cite{yoneta2024,chiba2024} that have subsequently been applied in holography to elegantly capture the microstates of single-sided black holes with smooth interiors \cite{wei2024,wei2024zez}.

Such structured TPQ states are constructed by imaginary-time evolving a crosscap state $|\mathcal{C}\rangle$ \cite{Ishibashi:1988kg,tu2017,caetano2022, Zhang:2022gtb,He:2023rrg,Tan:2024dcd,zhang2024,chalas2024,Li:2025sfn} as  $|\Psi_{\beta}\rangle\propto e^{-\beta H/4} |\mathcal{C}\rangle$ \cite{chiba2024, wei2024, yoneta2024}. Motivated by the fact that the bipartite entanglement remains constant under chaotic quenches from crosscap states \cite{chalas2024, wei2024}, in this Letter, we explore the quench dynamics of the bipartite entanglement entropy in the non-chaotic XX spin chain from the state $|\Psi_{\beta}\rangle$. We provide an exact analytical solution to this problem, where both the state preparation and the time evolution are governed by the same Hamiltonian $H$.

We study this problem from three complementary perspectives. First, using 2D conformal field theory (CFT), we derive an analytical formula for the time-dependent entanglement entropy in terms of Jacobi theta functions by evaluating vertex-operator correlators on the \textit{Klein bottle} geometry. Second, we obtain an exact numerical benchmark by solving the governing \textit{matrix Riccati equation} for the TPQ covariance matrix and evolving it in real time. Third, we show that the entire profile is captured by a quasiparticle picture that becomes quantitatively accurate in the scaling regime \cite{calabrese2005,Murciano:2021,Santalla:2022,Rottoli:2025}, where the ballistic propagation of antipodally entangled quasiparticle pairs governs the entropy double-plateau behavior.

{\em Thermal Pure Quantum State}. In 2D CFT, a crosscap state $|\mathcal{C}\rangle$ is a boundary state \cite{cardy1989npb} embodying a non-orientable spacetime, defined by the constraint $(L_n - (-1)^n \bar{L}_{-n})|\mathcal{C}\rangle = 0$ on the Virasoro generators \cite{Ishibashi:1988kg}. While abstract, its physical essence is captured on a lattice of length $L$ as an entangled antipodal pair (EAP) state \cite{chiba2024}. For a spin-1/2 chain, this is
\begin{equation}\label{crosscap}
    \ket{\mathcal{C}} = \bigotimes_{i=1}^{L/2} \frac{1}{\sqrt{2}} \left( \ket{\uparrow}_i \ket{\uparrow}_{i+L/2} + \ket{\downarrow}_i \ket{\downarrow}_{i+L/2} \right)\,.   
\end{equation}
This construction embeds maximal, non-local entanglement into the system, profoundly distinguishing it from short-range correlated ground states of local Hamiltonians.

A defining feature of $|\mathcal{C}\rangle$ [Eq.~(\ref{crosscap})] is its entanglement profile. The reduced density matrix $\rho_A$ of any contiguous subsystem $A$ of length $l \le L/2$ is maximally mixed, $\rho_A = \frac{I_A}{2^l}$. This results in a volume-law entanglement entropy, $S_A(0)=l\log 2$, with a full ``Page curve" profile $S_A(0) = \min(l, L-l) \log 2$. This property establishes the crosscap state as an infinite-temperature ($\beta=0$) TPQ state, as it perfectly mimics a thermal Gibbs state for any local operator.

We generalize this initial condition to a finite temperature by defining a generic TPQ state $|\Psi_{\beta}\rangle$ via imaginary time evolution:
\begin{equation}\label{TPQ}
    \ket{\Psi_{\beta}} \equiv \frac{e^{-\frac{\beta}{4}H} \ket{\mathcal{C}}}{\sqrt{\bra{\mathcal{C}}e^{-\frac{\beta}{2}H}\ket{\mathcal{C}}}}\,.
\end{equation}
This state is a deterministic, structured pure state which is, by construction, locally indistinguishable from a canonical Gibbs ensemble at inverse temperature $\beta$ \cite{yoneta2024}. The quench dynamics from $|\Psi_{\beta}\rangle$ thus explore the evolution from a state of ``fake" thermal equilibrium, allowing us to investigate how its globally stored quantum information is scrambled and redistributed over time.

{\em The CFT Approach}. The physical system under investigation is described at low energies by the $c=1$ free compact boson CFT, which is dual to the massless Dirac fermion. We now outline the analytical calculation of the time-dependent entanglement entropy following the quench from the state $|\Psi_{\beta}\rangle$. The system is prepared on a Euclidean cylinder of circumference $2\pi$ and length $\beta/2$ with crosscap boundary conditions, defining the TPQ state at $t=0$. This state then evolves in real time $t$ under the massless Hamiltonian $H$. We introduce the complex coordinates $(y, \bar{y})$ for this spacetime:  $y = \tau - i\sigma, \quad \bar{y} = \tau + i\sigma, \quad (0 \le \tau \le \frac{\beta}{2}, 0 \le \sigma \le 2\pi)$.

To compute the von Neumann entropy of a subsystem $A$, $S_A(t)$, we employ the replica trick, requiring the analytic continuation of the moments $\text{Tr}[\rho_A(t)^n]$ to $n \to 1$, where $\rho_A(t) =\text{Tr}_{\bar A}(e^{-iHt}|\Psi_\beta\rangle\langle\Psi_{\beta}|e^{iHt})$. The quantity $\text{Tr}[\rho_A(t)^n]$ is computed as a path integral on an $n$-sheeted Riemann surface representing the replicated spacetime. This problem can be mapped to a system of $n$ free Dirac fields on a single cylinder, where the replicas are coupled by twisted boundary conditions at the subsystem's endpoints \cite{Calabrese:2005}.

In the bosonized language, these fermionic replica twists are implemented by the insertion of specific twist fields, $\mathcal{T}_k$ and $\bar{\mathcal{T}}_k$, where $k$ labels the replica mode after a Fourier transform in the replica index. For the massless Dirac fermion, or equivalently for the compact boson at the free-fermion radius ($R=1$), these fermionic replica twist fields are primary vertex operators $V_{(\pm k, \pm k)}(y,\bar{y})=:e^{\pm i\frac{k}{n}(\varphi(y) + \bar{\varphi}(\bar{y}))}:$. The crucial insight is that the geometry of the initial state preparation (the crosscap) effectively transforms the problem of computing the replicated partition function to that of computing a two-point function of these vertex operators on a Klein bottle $\mathcal{K}$. The total replicated partition function is then the product over all replica modes:
\begin{equation}\label{TrrhoAn}
    \text{Tr}[\rho_A(t)^n] = \prod_{k=-\frac{n-1}{2}}^{\frac{n-1}{2}} Z_k = \prod_{k=-\frac{n-1}{2}}^{\frac{n-1}{2}} \braket{\mathcal{T}_k(y_1)\bar{\mathcal{T}}_k(y_2)}_{\mathcal{K}}\,.
\end{equation}
Here, $y_1$ and $y_2$ are the complex coordinates of the subsystem's endpoints in the spacetime describing the quench, given by $(y_1, \bar{y}_1) = (\frac{\beta}{4}+it -i\sigma_1, \frac{\beta}{4}+it + i\sigma_1)$ and $(y_2, \bar{y}_2) = (\frac{\beta}{4}+it - i\sigma_2, \frac{\beta}{4}+it + i\sigma_2)$. We denote by $\sigma=\sigma_2-\sigma_1$ the length of the subsystem.

In Eq.~(\ref{TrrhoAn}), we need to compute the normalized two-point function of vertex operators on a Klein bottle, which is a highly nontrivial task, although similar calculations have been carried out for the conformal boundary state in Ref.~\cite{Takayanagi:2010wp}. For a subsystem of length $\sigma$ at time $t$ after the quench from the TPQ state prepared with parameter $\beta$, the entanglement entropy is given by 
\begin{equation}\label{EntCFT}
    S_A(t, \sigma) = \frac{1}{6}\log\frac{\eta(\frac{i\beta}{2\pi})^{-6}
        \big|\theta_1(\frac{\sigma}{2\pi}|\frac{i\beta}{2\pi})\theta_2(\frac{\beta+4it}{4\pi i}|\frac{i\beta}{2\pi})\big|^2
    }{
         \big|\theta_2(\frac{\beta+4it+2i\sigma}{4\pi i}|\frac{i\beta}{2\pi})\theta_2(\frac{\beta+4it-2i\sigma}{4\pi i}|\frac{i\beta}{2\pi})\big|
    }\,.
\end{equation}
See the Supplemental Material~\cite{SM} for the derivation of this formula. Here and in the Supplemental Material we follow the string-theory conventions of Ref.~\cite{Polchinski:1998} with $\alpha'=2$. This expression involving the Dedekind eta function $\eta(\tau)$ and Jacobi theta functions $\theta_{1,2}(z|\tau)$, provides a complete, exact prediction for the entanglement dynamics, revealing the double-plateau structure and its dependence on $\beta$. Throughout this work, we adopt the conventions for the theta and eta functions as given in Ref. \cite{Polchinski:1998}. The derivation uses the vertex-operator representation of fermionic replica
twist fields and applies only to the free Dirac fermion. We do not claim that Eq.~(\ref{EntCFT}) holds for a generic compact boson at arbitrary radius.

{\em The Numerical Benchmark}. To validate our analytical CFT predictions, we perform exact numerical simulations on the lattice realization of the theory, namely the spin-1/2 XX chain at half filling. The Hamiltonian for a system of $L=2N$ sites with periodic boundary conditions is
\begin{equation}
    H = -J \sum_{j=1}^{2N} (\sigma_j^x \sigma_{j+1}^x + \sigma_j^y \sigma_{j+1}^y)\,.
\end{equation}
We take $J>0$ throughout. After a Jordan-Wigner transformation, $\sigma_j^+ = \left(\prod_{k=1}^{j-1}(-\sigma_k^z)\right)c_j^\dagger$, $\sigma_j^-= \left( \prod_{k=1}^{j-1} (-\sigma_k^z) \right) c_j, \sigma_j^z = 2c_j^\dagger c_j - 1$. The model is described by a Hamiltonian of non-interacting fermions, $H = -2J \sum_{j=1}^{2N} (c_j^\dagger c_{j+1} + \text{h.c.})$, with an anti-periodic boundary condition. The single-particle energy dispersion is given by $E(k) = -4J \cos(k)$.
The ground state is formed by filling all negative energy states, which corresponds to half-filling. The low-energy excitations occur near the two Fermi points, $k_F = \pm \pi/2$, where the dispersion becomes linear: $E(k_F \pm q) \approx \pm (4J)q$. This linear dispersion is the defining characteristic of the massless (1+1)D Dirac fermion, with an emergent ``speed of light" (Fermi velocity) of $v_F = 4J$ (in units where the lattice spacing $a=1$).

The initial TPQ state $|\Psi_{\beta}\rangle$ is prepared by evolving the crosscap state $|\mathcal{C}\rangle$ in imaginary time, as described in Eq.~(\ref{TPQ}). For the numerical implementation, we employ a fermionic representation of the system via the Jordan-Wigner transformation. It is important to note a subtle but important technical point regarding the lattice crosscap state itself. As was recently pointed out in Ref. \cite{zhang2024}, the Jordan-Wigner transformation of the strict spin crosscap state of Eq.~(\ref{crosscap}) results in a superposition of two distinct fermionic Gaussian states. For simplicity and to maintain consistency with the Gaussian state formalism used in our analytical and quasiparticle pictures, our numerical simulation, following the approach in Ref. \cite{chalas2024}, targets the evolution from a Gaussian version of the crosscap state $|\mathcal{C}\rangle=\bigotimes_{j=1}^{L/2} \frac{1}{\sqrt{2}}(1+c_j^{\dagger}c_{j+L/2}^{\dagger})|0\rangle$. We expect that this simplification does not affect the qualitative features of the entanglement dynamics reported here, as the essential non-local entanglement structure is captured by either Gaussian component, although a quantitative investigation of the full superposition remains an interesting direction for future work.

This setting allows us to leverage the powerful formalism of Gaussian states, as the entire quench protocol can be described exactly through the evolution of a fermionic covariance matrix $\Gamma$, which is defined in terms of Majorana fermion operators. For a system with $2N$ sites, we have $4N$ Majorana operators. The operators for site $j$ are $\gamma_{2j-1} = c_j + c_j^\dagger, \, \gamma_{2j} = i(c_j^\dagger - c_j)$, and the elements of fermionic covariance matrix $\Gamma$ are defined as $\Gamma_{mn} = \frac{i}{2} \langle [\gamma_m, \gamma_n] \rangle=i \langle \gamma_m \gamma_n \rangle-i\delta_{mn}$.

The initial crosscap state $\ket{\mathcal{C}}$ is a fermionic Gaussian state, and its properties are fully determined by its two-point correlation matrices. The normal correlator is $C_{ij}(0)\equiv\bra{\mathcal{C}}c_i^{\dagger}c_j\ket{\mathcal{C}} = \frac{1}{2}\delta_{ij}$, and the anomalous correlator is $F_{ij}(0)\equiv\bra{\mathcal{C}}c_ic_j\ket{\mathcal{C}} = \frac{1}{2}(\delta_{j, i+N} - \delta_{i, j+N})$. It is more convenient to combine these into the $4N \times 4N$ real, anti-symmetric fermionic covariance matrix, $\Gamma_{\mathcal{C}}$. For the crosscap state, $\Gamma_{\mathcal{C}}$ is zero except for $2 \times 2$ blocks connecting antipodal sites $i$ and $i+N$, which take the form $\Gamma_{i,i+N} = \sigma_x$.

The imaginary time evolution operator $e^{-\beta H/4}$ is a Gaussian operator, meaning the initial TPQ state $\ket{\Psi_\beta}$ is also a Gaussian state. Its covariance matrix, $\Gamma_\beta$, can be found by solving the governing \textit{matrix Riccati equation} \cite{AbouKandil:2003,Bellman:1978,Paviglianiti:2024} for the imaginary time evolution \cite{kraus2010,Ashida:2018}.
\begin{equation}
    \frac{d\Gamma(\tau)}{d\tau} = -\mathcal{H} - \Gamma(\tau)\mathcal{H}\Gamma(\tau)\,.
    \label{eq:riccati}
\end{equation}
The equation above holds if the Hamiltonian is quadratic: $H =\sum_{i,j=1}^{2N} h_{ij} c_i^\dagger c_j= \frac{i}{4} \sum_{ij} \mathcal{H}_{ij} \gamma_i \gamma_j$. By applying a trick, we can linearize this equation and solve it exactly as   
\begin{equation}\label{Gammatau}
    \Gamma(\tau) =(\cos(\mathcal{H}\tau)\Gamma_0 - \sin(\mathcal{H}\tau)) (\sin(\mathcal{H}\tau)\Gamma_0+\cos(\mathcal{H}\tau))^{-1}\,,
\end{equation}
where $\mathcal{H}$ is the $4N \times 4N$ matrix representation of the single-particle Hamiltonian in the Majorana basis. The derivation of Eq.~(\ref{eq:riccati}) and Eq.~(\ref{Gammatau}), together with the imaginary-time and real-time covariance-matrix evolution are given in the Supplemental Material~\cite{SM}. For the XX spin chain, $\mathcal{H}= \mathbf{h} \otimes (i\sigma_y)$. The covariance matrix for the TPQ state is then obtained by taking $\tau=\frac{\beta}{4}$ in Eq.~(\ref{Gammatau}), \textit{i.e.} $\Gamma_{\beta}=\Gamma(\tau=\frac{\beta}{4})$, with $\Gamma_0=\Gamma_{\mathcal{C}}$.

Having obtained the exact covariance matrix $\Gamma_\beta$ for the initial state, the real-time evolution for the quench is given by the following equation:
\begin{equation}
    \Gamma(t) = e^{\mathcal{H}t} \Gamma_\beta e^{-\mathcal{H}t}\,.
\end{equation}
The entanglement entropy of a subsystem $A$ of length $l$ at any time $t$ can then be computed directly from the eigenvalues $\pm i\nu_j$ of its restricted covariance matrix $\Gamma_A(t)$, which is the $2l \times 2l$ sub-block of $\Gamma(t)$ corresponding to the sites in $A$. The von Neumann entropy is a sum of binary entropies:
\begin{equation}
    S_A(t) = -\sum_{j=1}^l \left[ \frac{1+\nu_j}{2}\ln\frac{1+\nu_j}{2} + \frac{1-\nu_j}{2}\ln\frac{1-\nu_j}{2} \right]\,.
\end{equation}
This procedure provides an exact, numerically efficient method to simulate the dynamics for large systems, serving as a rigorous benchmark for our analytical CFT results.
\begin{figure}
        \centering
        {\includegraphics[width=4.2cm]{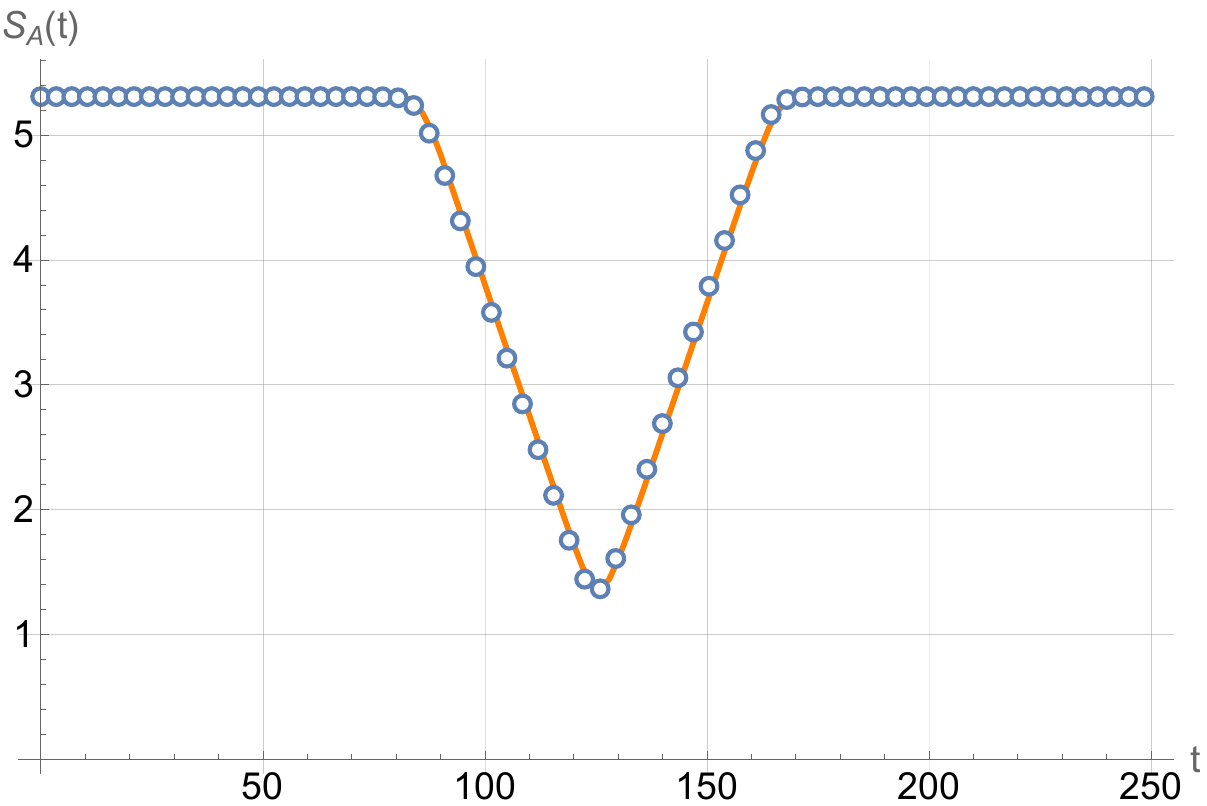}}
        {\includegraphics[width=4.2cm]{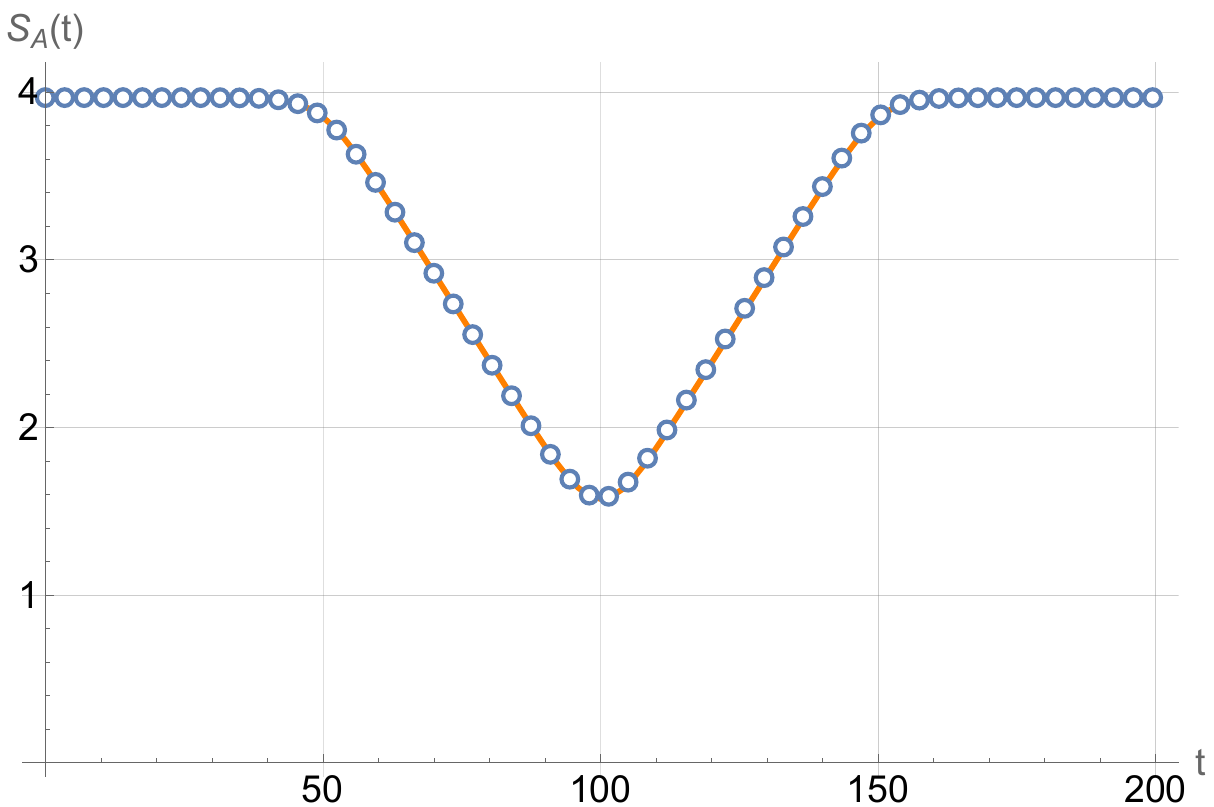}}
        \caption{Comparison between the exact numerical results and the CFT prediction for the entanglement entropy $S_A(t)$. Solid lines: numerical data; circles: CFT prediction. Left: $L=500, l=80, \beta =20, v_F=1 (J=0.25)$. Right: $L=400,l=100,\beta=40, v_F=1 (J=0.25)$.}
        \label{fig1}
\end{figure}

As shown in Fig.~{\ref{fig1}}, the CFT prediction (symbols) is in excellent quantitative agreement with the exact lattice calculation (solid line), accurately capturing both the double-plateau structure and the characteristic time scales after the rescaling $\beta\to 2\pi v_F\beta/L, \sigma\to 2\pi l/L, t\to 2\pi v_Ft/L$ in Eq.~(\ref{EntCFT}).

{\em The Quasiparticle Picture}. The entanglement dynamics can be quantitatively understood through the quasiparticle picture \cite{calabrese2005,Murciano:2021,Santalla:2022,Rottoli:2025}. In the standard quench paradigm from a low-entanglement state, the quench acts as a source of locally entangled quasiparticle pairs that propagate ballistically, causing entanglement to grow. The TPQ state, however, is a high-energy, extensively entangled state whose dynamics require a fundamentally different interpretation. Here, the dominant process is not the creation of entanglement, but the transport and eventual destruction of the pre-existing, non-local entanglement.

The initial state $|\Psi_\beta\rangle$ can be viewed as a sea of quasiparticle pairs with opposite momenta $(k, -k)$. Crucially, the entanglement is non-local: a quasiparticle with momentum $k$ at site $x$ is entangled not with its partner at $x$, but with the particle of momentum $-k$ at the antipodal site $x+L/2$. After the quench at $t=0$, these pairs propagate ballistically. The entanglement entropy of a subsystem $A$ decreases when both members of an entangled pair become fully contained within $A$. This process leads to the following prediction for the entanglement dynamics:
\begin{equation}
    S_A(t) = S_A(0) - 2 \int_{-\pi}^{\pi} \frac{dk}{2\pi} s(k) \max\left(0, l-|2v(k)\tau_k-L/2|\right)\,,
    \label{eq:qp_formula}
\end{equation}
which is a straightforward generalization of the formula proposed in Ref. \cite{chalas2024}. We now explain this formula term by term. The first term, $S_A(0)=l\int_{-\pi}^{\pi} \frac{dk}{2\pi} s(k)$, is the initial entanglement entropy of the subsystem of length $l$. The second term is a subtractive correction representing the entropy removed by the quasiparticles. The factor of 2 accounts for the two quasiparticles in a pair. The counting function, $\max(0, \dots)$, is the heart of the picture: it counts the number of antipodal pairs with momentum $k$ that are both inside the subsystem $A$ at time $t$. The term $|2v(k)\tau_k - L/2|$ is the effective separation of the two members of an antipodal pair, which starts at $L/2$ and decreases as they travel towards each other. Here, $v(k)$ is the velocity of a quasiparticle with momentum $k$. For the XX chain at half filling, $v(k) =4J \sin(k)$ and $\tau_k = t \pmod{L/|v(k)|}$ is a periodic time variable that accounts for revivals in the finite system.

The function $s(k)$ is the effective entanglement density carried by each quasiparticle with momentum $k$. As shown in the Supplemental Material~\cite{SM}, the initial state is equivalent to a BCS-like state 
\begin{equation}\label{TPQBCS}
|\Psi_{\beta}\rangle = \prod_{k>0} \left( u_k(\beta) + v_k(\beta) c_k^\dagger c_{-k}^\dagger \right) |0\rangle
\end{equation}
with Bogoliubov coefficients
\begin{equation}
\begin{split}
|u_k(\beta)|^2 = \frac{1}{1+e^{-\beta E(k)}},\qquad |v_k(\beta)|^2 =  \frac{e^{-\beta E(k)}}{1+e^{-\beta E(k)}}\,. 
\end{split}
\end{equation}
This allows for a direct calculation of the entanglement entropy per mode, which is given by the binary entropy of the mode occupation:
\begin{equation}
    s(k) = -n_k\ln n_k - (1-n_k)\ln(1-n_k)\,,
\end{equation}
where the occupation $n_k$ is given by the Fermi-Dirac distribution at the inverse temperature $\beta$ of the initial state:
\begin{equation}
    n_k\equiv\bra{\Psi_\beta}c_k^{\dagger}c_k\ket{\Psi_\beta}= \frac{1}{e^{\beta E(k)} + 1}\,.
\end{equation}
\begin{figure}
        \centering
        {\includegraphics[width=4.2cm]{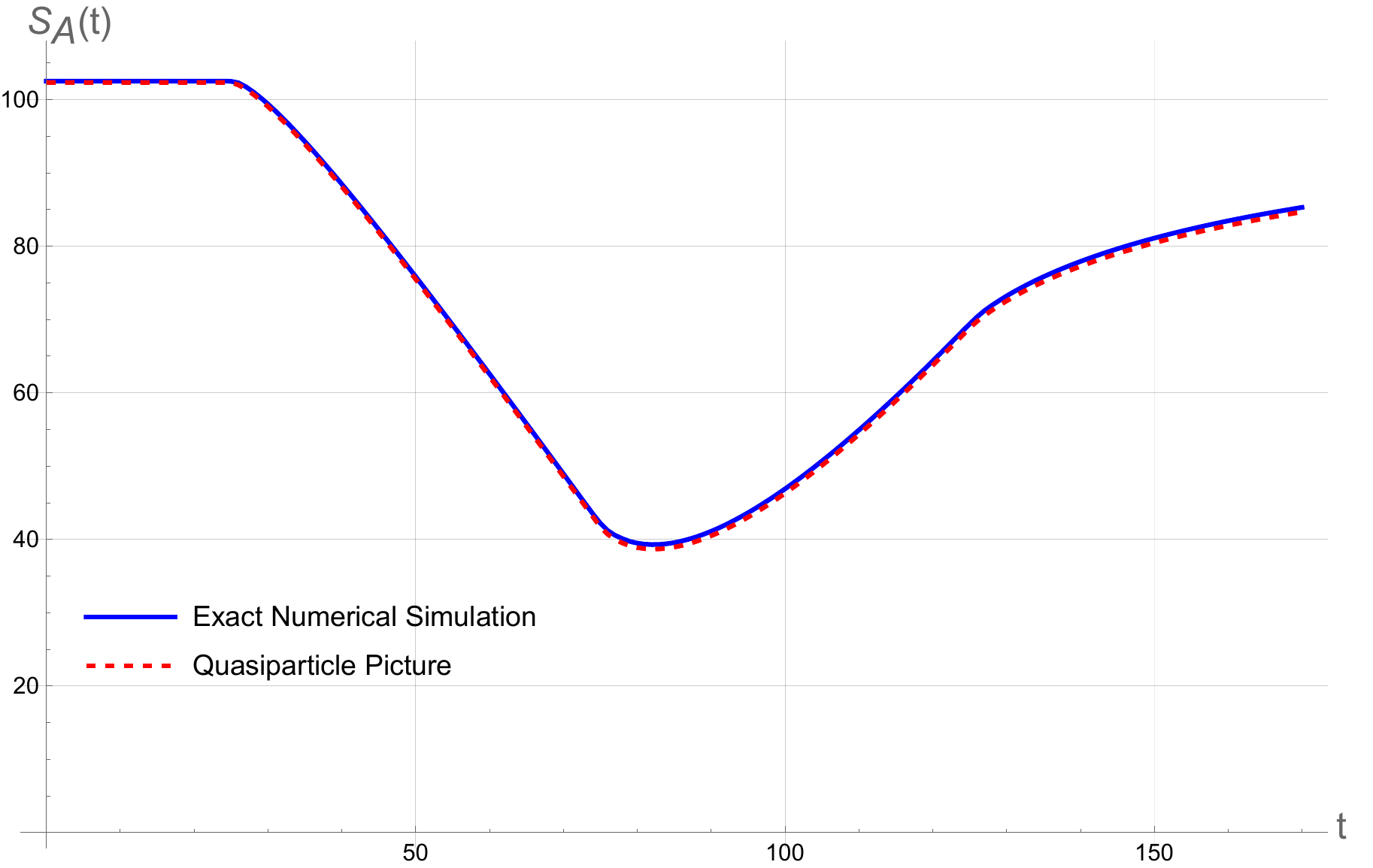}}
        {\includegraphics[width=4.2cm]{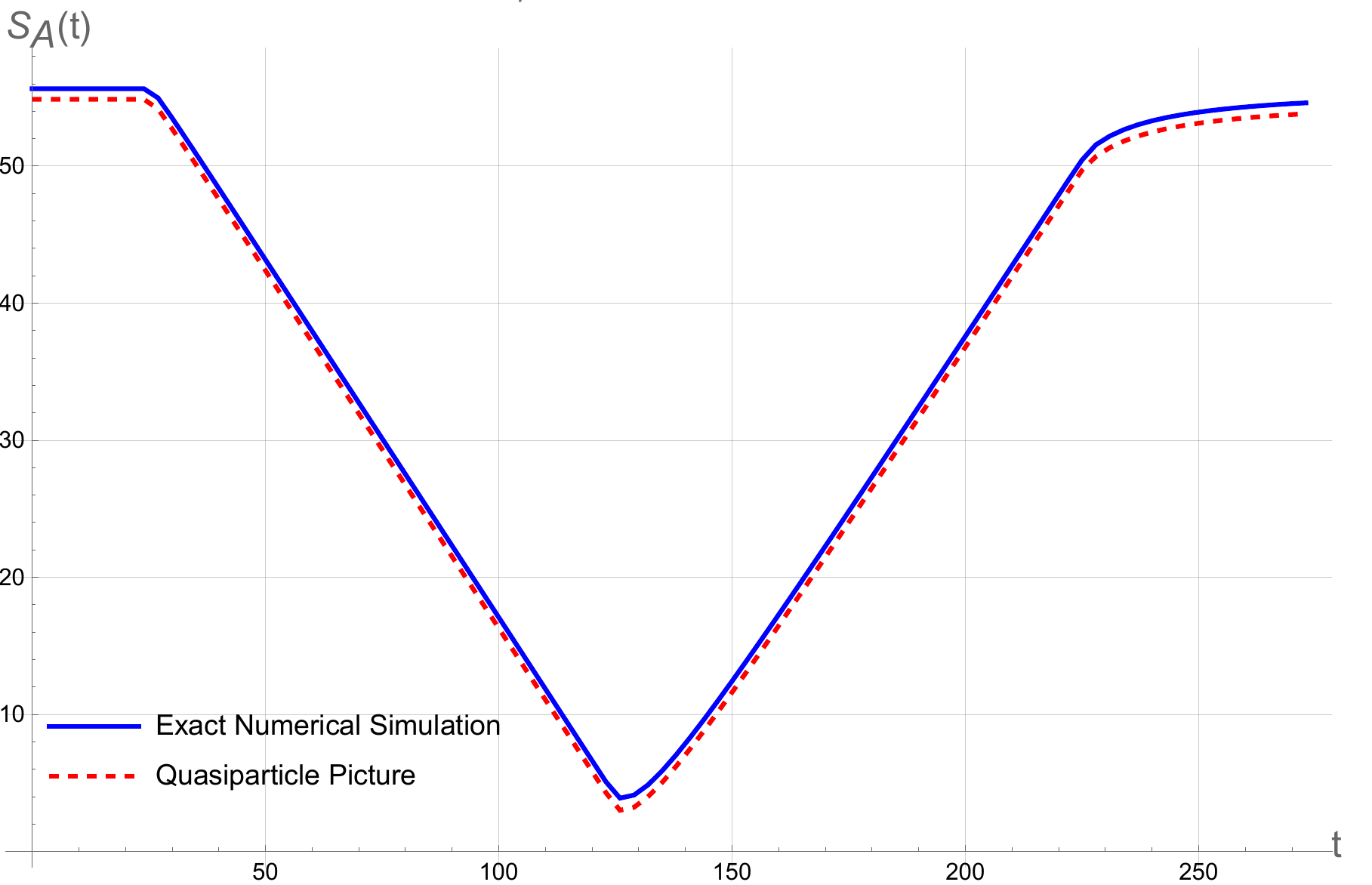}}
        \caption{Comparison between the exact numerical results and the quasiparticle prediction for the entanglement entropy $S_A(t)$. Solid lines: numerical data; dashed line: quasiparticle prediction. Left: $L=600, l=200, \beta=1, J=0.5$. Right: $L=1000,l=400,\beta=4,J=0.5$.}
        \label{fig2}
\end{figure}
The formula (\ref{eq:qp_formula}) provides a complete quantitative picture that accurately reproduces our exact analytical and numerical results. The initial plateau is the time before any antipodal pairs can fully enter the subsystem. The subsequent decrease is governed by the rate at which pairs enter, weighted by their entanglement density $s(k)$. The rise and second plateau are due to the same pairs exiting the subsystem as they traverse outside the subsystem. This quasiparticle picture thus provides a powerful physical interpretation for the non-monotonic behavior of entanglement observed in the system. Figure~\ref{fig2} shows a comparison between the exact numerical results and the quasiparticle prediction for the entanglement entropy.

{\em Conclusion and outlook}. In this work, we study the entanglement evolution following a quench from a volume-law entangled TPQ state. We establish this exotic entanglement dynamics rigorously through three complementary approaches: (i) a 2D CFT derivation yielding an exact analytical formula in terms of Jacobi theta functions on the Klein bottle geometry, (ii) a highly efficient exact numerical evaluation of the restricted covariance matrix, and (iii) a quasiparticle picture that becomes quantitatively accurate in the scaling regime. 

Our results pave the way for exploring nonequilibrium dynamics from highly entangled initial states and for extending the CFT quench framework to non-orientable manifolds. The CFT analysis and exact Gaussian numerics apply directly to mutual information and entanglement negativity, while the quasiparticle picture can be adapted with minor modifications. Exploring interacting integrable models in this context would be both interesting and important for future work. This perspective can be generalized to richer scenarios, including inhomogeneous quenches such as the Möbius or sine-square deformation (SSD) \cite{Okunishi:2016,Wen:2018,Kawano:2022,Nozaki:2024,Bernamonti:2024,Hotta:2021}.
The experimental exploration of crosscap-type initial states and the associated entanglement dynamics would be a fascinating direction for future studies, potentially realizable in quantum-simulation platforms such as ultracold atoms or superconducting qubits.

{\em Acknowledgments}. I am grateful to Pasquale Calabrese and Konstantinos Chalas for very helpful discussions. This work was supported by the National Natural Science Foundation of China (Grant Nos. 12005081, and 12465014), and the program of the China Scholarship Council (Grant No. 202308360098).
\bibliographystyle{apsrev4-1}

\clearpage
\onecolumngrid
\begin{center}
{\large \textbf{Supplemental Material for\\
Exact Quench Dynamics from Thermal Pure Quantum States}}

\vspace{0.5cm}

Hui-Huang Chen
\end{center}

\setcounter{equation}{0}
\setcounter{figure}{0}
\setcounter{table}{0}
\setcounter{section}{0}

\renewcommand{\theequation}{S\arabic{equation}}
\renewcommand{\thefigure}{S\arabic{figure}}
\renewcommand{\thetable}{S\arabic{table}}
\renewcommand{\thesection}{S\Roman{section}}

\tableofcontents

\section{The CFT approach to the quench dynamics from TPQ states}

In this section, we give the full derivation of formula (4) in the main text. 
\subsection{Crosscap state in free compact boson CFT}
We work in the bosonized description of the free massless Dirac fermion, setting \(\alpha' = 2\) in string theory conventions \cite{Polchinski:1998}. 
We consider a free scalar field $\phi(y,\bar y)=\varphi(y)+\bar\varphi(\bar y)$ on the cylinder with circumference $2\pi$. The field is compactified on a circle of radius $R$, satisfying
\begin{equation} 
\phi(\sigma+2\pi,t)=\phi(\sigma,t)+2\pi Rw\,,
\end{equation}
where $w\in\mathbb{Z}$ is the winding number. We can expand holomorphic part and anti-holomorphic part of the field as
\begin{equation}
\varphi(y) = \varphi_0 - i \alpha_0 y + i \sum_{m \neq 0} \frac{\alpha_m}{m} e^{-my}, \quad  
\bar\varphi(\bar y) = \bar\varphi_0 - i \bar\alpha_0 \bar{y} + i \sum_{m \neq 0} \frac{\bar{\alpha}_m}{m} e^{-m\bar{y}}\,,
\end{equation}
where we have defined
\begin{equation}\label{zeromode}
\alpha_0=\frac{n}{R}+\frac{wR}{2},\qquad \bar\alpha_0=\frac{n}{R}-\frac{wR}{2}
\end{equation}
with commutation relations:
\begin{equation}
[\varphi_0, \alpha_0] = i\,, \quad [\bar\varphi_0,\bar\alpha_0]=i\,, \quad [\alpha_m, \alpha_n] = m \delta_{n+m,0}\,, \quad [\bar\alpha_m, \bar\alpha_n] = m \delta_{n+m,0}\,.
\end{equation}
The rest of the commutators vanish. 

The Hamiltonian is $H = L_0 +\bar{L}_0-\frac{1}{12}$, with
\begin{equation} 
L_0=\frac12\big(\frac{n}{R}+\frac{wR}{2}\big)^2+\sum_{n=1}^{\infty}\alpha_{-n}\alpha_n\,,\qquad \bar L_0=\frac12\big(\frac{n}{R}-\frac{wR}{2}\big)^2+\sum_{n=1}^{\infty}\bar\alpha_{-n}\bar\alpha_n\,.
\end{equation}
The vacuum state is labeled by the winding number $w$ and the momentum integer $n$ and is denoted by $\ket{n,w}$.
\par The crosscap state satisfies
\begin{equation}\label{constraint} 
(\alpha_n-(-1)^n\bar\alpha_{-n})\ket{\mathcal{C}}=0\,.
\end{equation}
The gluing condition above also fixes the zero-mode sector. Setting $n=0$ in
Eq.~(\ref{constraint}) gives
\begin{equation}
(\alpha_0-\bar\alpha_0)|\mathcal{C}\rangle=0.
\end{equation}
Using Eq.~(\ref{zeromode}), this condition becomes
\begin{equation}
wR\,|\mathcal{C}\rangle=0,
\end{equation}
and therefore only states with vanishing winding number, $w=0$, can
contribute to the zero-mode part of the crosscap state. The momentum
quantum number remains unrestricted. Thus the zero-mode part is
proportional to
\begin{equation}
\sum_{n\in\mathbb Z}|n,0\rangle .
\end{equation}
Combining this zero-mode contribution with the oscillator solution gives
\begin{equation}\label{BN}
\ket{\mathcal{C}} = \mathcal{N} e^{\sum_{n=1}^\infty \frac{(-1)^n}{n} \alpha_{-n} \bar{\alpha}_{-n}}\sum_{n=-\infty}^\infty\ket{n,0}\,,
\end{equation}
where \(\mathcal{N}\) is a normalization factor.

Although the subsequent calculations largely parallel those in \cite{Takayanagi:2010wp}, subtle differences arise at certain points, ultimately resulting in a qualitatively distinct behavior of the entanglement entropy.
\subsection{The two-point functions}
The unnormalized two-point function of the vertex operators $V_{k,\bar k} = :e^{i k\varphi + i\bar k\bar\varphi}:$ on the cylinder is
\begin{equation}\label{corr}
\bra{\mathcal{C}}e^{-\beta H/2} V_{k, \bar k}(y_1, \bar{y}_1) V_{-k, -\bar k}(y_2, \bar{y}_2)\ket{\mathcal{C}}\,.
\end{equation}
Using the Baker-Campbell-Hausdorff (BCH) formula, $e^A e^B = e^{A+B+\frac{1}{2}[A,B]}$, the zero-mode part of the correlator above evaluates to
\begin{equation}\label{BCH}
\sum_{n=-\infty}^\infty e^{-\frac{n^2\beta}{2R^2}} e^{\frac{n}{R}\left(k (y_1-y_2) + \bar k (\bar{y}_1-\bar{y}_2) \right)} e^{\frac{k^2}{2}(y_2-y_1) + \frac{\bar k^2}{2}(\bar{y}_2-\bar{y}_1)}\,.
\end{equation} 
\par To deal with the massive modes, let us first focus on a single mode, say the $n$-th massive mode. Notice that $[\alpha_n,\alpha_{-n}]=n$,
which means if we define $\hat{\alpha}\equiv\frac{e^{\frac{\pi i n}{2}}\alpha_n}{\sqrt{n}}$ and $\hat{\alpha}^{\dagger}\equiv\frac{e^{-\frac{\pi i n}{2}}\alpha_{-n}}{\sqrt{n}}$, then $\hat{\alpha}$ and $\hat{\alpha}^{\dagger}$ are the standard harmonic creation and annihilation operators, satisfying $[\hat{\alpha},\hat{\alpha}^{\dagger}]=1$. Similarly, for the anti-holomorphic part, we define $\hat{\beta}\equiv\frac{e^{\frac{\pi i n}{2}}\bar\alpha_n}{\sqrt{n}}$ and $\hat{\beta}^{\dagger}\equiv\frac{e^{-\frac{\pi i n}{2}}\bar\alpha_{-n}}{\sqrt{n}}$, which also satisfy $[\hat{\beta},\hat{\beta}^{\dagger}]=1$. These oscillators are contained in  
\begin{equation} 
\begin{split}
&ik\varphi(y_i):~ -\frac{ke^{-\frac{\pi i n}{2}}}{\sqrt{n}}e^{-ny_i}\hat{\alpha}+\frac{ke^{\frac{\pi i n}{2}}}{\sqrt{n}}e^{ny_i}\hat{\alpha}^{\dagger},\qquad \text{with}~ n>0\\
&i\bar k\bar\varphi(y_i):~ -\frac{\bar ke^{-\frac{\pi i n}{2}}}{\sqrt{n}}e^{-n\bar y_i}\hat{\beta}+\frac{\bar ke^{\frac{\pi i n}{2}}}{\sqrt{n}}e^{n\bar y_i}\hat{\beta}^{\dagger},\qquad \text{with}~ n>0\,.
\end{split}
\end{equation}
Then $V_{k, \bar k}(y_1, \bar{y}_1) V_{-k, -\bar k}(y_2, \bar{y}_2)$ contains $ e^{a\hat{\alpha}+\bar a\hat{\beta}} e^{b\hat{\alpha}^\dagger+\bar b\hat{\beta}^\dagger}$, where
\begin{equation}
\begin{split}  
&a = -\frac{e^{-\frac{\pi i n}{2}}k}{\sqrt{n}} \left( e^{-n y_1} - e^{-n y_2} \right), \quad  
\bar a = -\frac{e^{-\frac{\pi i n}{2}}\bar k}{\sqrt{n}} \left( e^{-n\bar{y}_1} - e^{-n\bar{y}_2} \right),  
\\
&b = \frac{e^{\frac{\pi i n}{2}}k}{\sqrt{n}} \left(e^{n y_1} - e^{n y_2}\right), \quad  
\bar b = \frac{e^{\frac{\pi i n}{2}}\bar k}{\sqrt{n}} \left(e^{n \bar{y}_1} - e^{n\bar{y}_2} \right)\,.
\end{split}
\end{equation}
In the definition of the crosscap state [Eq.~(\ref{BN})], the factor is $e^{\hat{\alpha}^{\dagger}\hat{\beta}^{\dagger}}$. They also appear in the leftmost of Eq.~(\ref{corr}): $\bra{0}e^{\hat{\alpha}\hat{\beta}}e^{-\frac{n\beta}{2}(\hat{\alpha}^{\dagger}\hat{\alpha}+\hat{\beta}^{\dagger}\hat{\beta})}=\bra{0}e^{z\hat{\alpha}\hat{\beta}}$, with $z=e^{-n\beta}$, where we have used the BCH formula $e^Xe^Y=\exp(e^{ad_X}Y)e^X$. Therefore, we need to compute $\bra{0}e^{\hat{\alpha}\hat{\beta} z} e^{a\hat{\alpha} + \bar\alpha\hat{\beta}} e^{b\hat{\alpha}^{\dagger} + \bar b\hat{\beta}^{\dagger}} e^{\hat{\alpha}^\dagger\hat{\beta}^\dagger}\ket{0}$, and the final result is
\begin{equation}\label{identity}
\bra{0}e^{\hat{\alpha} \hat{\beta} z} e^{a\hat{\alpha} + \bar\alpha\hat{\beta}} e^{b\hat{\alpha}^\dagger + \bar b\hat{\beta}^\dagger} e^{\hat{\alpha}^\dagger\hat{\beta}^\dagger}\ket{0}= \frac{1}{1-z}e^{\frac{ab + \bar a\bar b+a\bar a+zb\bar b}{1-z}}\,.
\end{equation}
This identity can be proved in several ways, and we present one proof at the end of this section.

We should sum over all the massive modes as $\prod_{n=1}^{\infty}\frac{1}{1-z}\prod_{m=0}^{\infty}\exp(z^m(ab + \bar a\bar b+a\bar a+zb\bar b))$.
Here we have used the identity $\frac{1}{1-z}=\sum_{m=0}^{\infty}z^m$. The constant term $-\frac{1}{12}$ in the Hamiltonian and the term $\prod_{n=1}^{\infty}\frac{1}{1-z}$ are directly related to Dedekind $\eta$-function
\begin{equation} 
e^{\frac{1}{12}\frac{\beta}{2}}\prod_{n=1}^{\infty}\frac{1}{1-z}=\frac{1}{\eta\left(\frac{i\beta}{2\pi}\right)}\,.
\end{equation}
The product of the two normal-ordered vertex operators has to be reordered before applying Eq.~(\ref{identity}). This reordering gives
an extra factor which cancels the $m=0$ contributions from the purely holomorphic and antiholomorphic terms $z^m ab$ and $z^m\bar a\bar b$.
Thus, focusing on the holomorphic term, we obtain
\begin{equation}\label{S14}
\prod_{n=1}^{\infty}
\exp\!\left[
\frac{k^2}{n}e^{n(y_2-y_1)}
\right]
\prod_{m=1}^{\infty}\prod_{n=1}^{\infty}
\exp\!\left[
\frac{k^2 e^{-mn\beta}}{n}
\left(
e^{n(y_2-y_1)}
+
e^{-n(y_2-y_1)}
-2
\right)
\right]
=
e^{-\frac{k^2}{2}(y_2-y_1)}
\left[
\frac{
\eta\!\left(\frac{i\beta}{2\pi}\right)^3
}{
\theta_1\!\left(
\frac{y_2-y_1}{2\pi i}
\,\middle|\,
\frac{i\beta}{2\pi}
\right)
}
\right]^{k^2}.
\end{equation}
Here we have used $\sum_{n=1}^{\infty}x^n/n=-\log(1-x)$ and the product definitions of the eta and theta functions~\cite{Polchinski:1998}. The
antiholomorphic contribution is obtained by replacing $k,y_i$ with $\bar k,\bar y_i$. 
\par The other terms can be computed in a similar way:
\begin{equation} 
\prod_{n=1}^{\infty}\prod_{m=0}^{\infty}\exp(z^m(a\bar a+zb\bar b))=\left(\frac{\theta_2\left(\frac{y_1+\bar{y}_1}{2\pi i}|\frac{i\beta}{2\pi}\right)\theta_2\left(\frac{y_2+\bar{y}_2}{2\pi i}|\frac{i\beta}{2\pi}\right)}{\theta_2\left(\frac{y_1+\bar{y}_2}{2\pi i}|\frac{i\beta}{2\pi}\right)\theta_2\left(\frac{y_2+\bar{y}_1}{2\pi i}|\frac{i\beta}{2\pi}\right)}\right)^{-k\bar k}\,.
\end{equation}
Combining all these terms, we get
\begin{equation}\label{twistcorrelator}
\begin{split}
&\langle \mathcal{C} | e^{-\frac{\beta}{2} H} V_{(k, \bar k)}(y_1, \bar{y}_1) V_{(-k, -\bar k)}(y_2, \bar{y}_2) |\mathcal{C} \rangle\\
&= \mathcal{N}^2 \left[ \sum_{n=-\infty}^\infty e^{-\frac{n^2\beta}{2R^2}} e^{\frac{n}{R}\left(k (y_1-y_2) + \bar k (\bar{y}_1-\bar{y}_2) \right)}\right] \cdot \frac{1}{\eta\left(\frac{i\beta}{2\pi}\right)} \\
&\left(\frac{\eta\left(\frac{i\beta}{2\pi}\right)^3}{\theta_1\left(\frac{y_2-y_1}{2\pi i}|\frac{i\beta}{2\pi}\right)}\right)^{k^2}
\left(\frac{\eta\left(\frac{i\beta}{2\pi}\right)^3}{\theta_1\left(\frac{\bar y_2-\bar y_1}{2\pi i}|\frac{i\beta}{2\pi}\right)}\right)^{\bar k^2}
\cdot \left(\frac{\theta_2\left(\frac{y_1+\bar{y}_1}{2\pi i}|\frac{i\beta}{2\pi}\right)\theta_2\left(\frac{y_2+\bar{y}_2}{2\pi i}|\frac{i\beta}{2\pi}\right)}{\theta_2\left(\frac{y_1+\bar{y}_2}{2\pi i}|\frac{i\beta}{2\pi}\right)\theta_2\left(\frac{y_2+\bar{y}_1}{2\pi i}|\frac{i\beta}{2\pi}\right)}\right)^{-k\bar k}\,.
\end{split}
\end{equation}
The exponential prefactors in Eq.~(\ref{S14}) and in its antiholomorphic counterpart cancel the corresponding zero-mode BCH factors in
Eq.~(\ref{BCH}). After substituting the explicit values of the coordinates 
$(y_1, \bar{y}_1) = (\frac{\beta}{4}+it -i\sigma_1, \frac{\beta}{4}+it + i\sigma_1)$ and $(y_2, \bar{y}_2) = (\frac{\beta}{4}+it - i\sigma_2, \frac{\beta}{4}+it + i\sigma_2)$ and taking into account that twist fields at $R=1$ having $k=\bar k$, the second line of the equation above simplifies as 
\begin{equation}\label{norm}
\frac{\sum_{n=-\infty}^\infty e^{-\frac{n^2\beta}{2}}}{\eta\left(\frac{i\beta}{2\pi}\right)}=\frac{\theta(0|\frac{i\beta}{2\pi})}{\eta\left(\frac{i\beta}{2\pi}\right)}\,.
\end{equation}
This is exactly equal to $\bra{\mathcal{C}}e^{-\frac{\beta}{2}H}\ket{\mathcal{C}}$, which can be interpreted as the partition function on the Klein bottle.
\subsection{The entanglement entropy}
To calculate the von Neumann entanglement entropy of a subsystem $A$, $S_A(t)$, we employ the replica trick. This requires first computing $\text{Tr}(\rho_A(t)^{N})$ and then performing an analytic continuation in the replica index $N \to 1$:
\begin{equation}\label{replica}
    S_A(t) = -\frac{\partial}{\partial N} \log\left( \text{Tr}[\rho_A(t)^N] \right) \Big|_{N=1}\,.
\end{equation}
The quantity $\text{Tr}[\rho_A(t)^N]$ is computed as a partition function on an $N$-sheeted Riemann surface. This problem can be mapped to a system of $N$ free Dirac fermion fields, $\{\psi^{(a)}\}_{a=1}^N$, on a single cylinder, where the interaction between replicas is encoded in twisted boundary conditions at the endpoints of the subsystem $A$. For example, at endpoint $y_1$, we have
\begin{equation}
    \psi^{(a)}(e^{2\pi i} y_1) = \psi^{(a+1)}(y_1)\,.
\end{equation}
This system of coupled fields can be diagonalized by a discrete Fourier transform in the replica index $a$, resulting in $N$ decoupled fermion theories, each with a diagonal phase twist at the boundaries. For the $k$-th mode, the boundary condition becomes
\begin{equation}
    \psi_k(e^{2\pi i} y_1) = e^{2\pi i k/N} \psi_k(y_1)\,.
\end{equation}
\par A (1+1)D free massless Dirac fermion, with left- and right-moving components $\psi, \bar\psi$, is equivalent to a free scalar boson $\phi$ compactified on a circle of radius $R=1$. The fermionic fields are represented as exponentials of the bosonic field:
\begin{equation}
    \psi(y) = e^{i\varphi(y)}, \quad \bar\psi(\bar{y}) = e^{i\varphi(\bar{y})}\,,
\end{equation}
where $\phi(y, \bar{y}) = \varphi(y) + \bar\varphi(\bar{y})$.

A twisted boundary condition for a fermion, such as $\psi_k(e^{2\pi i} y_1) = e^{2\pi i k/N} \psi_k(y_1)$, is equivalent to the insertion of a specific local operator, a twist field, at that point.
In the bosonized theory, the twist fields that implement the fermionic replica twists are vertex operators. For the $k$-th decoupled theory, the boundary condition is created by inserting a pair of twisted-sector vertex operators, $\mathcal{T}_{k}$ and $\bar{\mathcal{T}}_{k}$, at the endpoints of the interval, $y_1$ and $y_2$. These vertex operators are primary fields of the bosonic CFT. For the Dirac fermion, they are
\begin{align}
    \mathcal{T}_{k}(y, \bar{y}) &= V_{(\frac{k}{N}, \frac{k}{N})}(y, \bar{y}) = :e^{i\frac{k}{N}(\varphi(y) + \bar{\varphi}(\bar{y}))}: \\
    \bar{\mathcal{T}}_{k}(y, \bar{y}) &= V_{(-\frac{k}{N}, -\frac{k}{N})}(y, \bar{y}) = :e^{-i\frac{k}{N}(\varphi(y) + \bar{\varphi}(\bar{y}))}:\,,
\end{align}
where $V_{(k, \bar k)}$ is the general vertex operator with left and right charges $(k, \bar k)$. 

With this mapping, the problem of calculating the partition function for the $k$-th replicated fermion is transformed into calculating a two-point function of bosonic vertex operators on a Klein bottle
\begin{equation}
    Z_k = \langle \mathcal{T}_{k}(y_1, \bar{y}_1) \bar{\mathcal{T}}_{k}(y_2, \bar{y}_2)\rangle_{\mathcal{K}}\,.
\end{equation}
The total replicated partition function is then the product over all modes $k$
\begin{equation}
    \text{Tr}[\rho_A(t)^N] = \prod_{k=-\frac{N-1}{2}}^{\frac{N-1}{2}} Z_k=\prod_{k=-\frac{N-1}{2}}^{\frac{N-1}{2}}\frac{\bra{\mathcal{C}}e^{-\frac{\beta}{2}H}\mathcal{T}_{k}(y_1, \bar{y}_1) \bar{\mathcal{T}}_{k}(y_2, \bar{y}_2)\ket{\mathcal{C}}}{\bra{\mathcal{C}}e^{-\frac{\beta}{2}H}\ket{\mathcal{C}}}\,.
    \label{eq:trace_as_correlator}
\end{equation}
The crucial ingredient for this calculation is the explicit form of the normalized two-point function on the Klein bottle. This has already been done in the previous section (cf. Eq.~(\ref{twistcorrelator}) and Eq.~(\ref{norm})), and the result is
\begin{equation}\label{twist2pt}
    \langle \mathcal{T}_{k}(y_1, \bar{y}_1) \bar{\mathcal{T}}_{k}(y_2, \bar{y}_2)\rangle_{\mathcal{K}} = \left( \frac{
        \eta(\frac{i\beta}{2\pi})^6 |\theta_2(\frac{\beta+4it}{4\pi i}+\frac{\sigma}{2\pi}|\frac{i\beta}{2\pi})||\theta_2(\frac{\beta+4it}{4\pi i}-\frac{\sigma}{2\pi}|\frac{i\beta}{2\pi})|
    }{
        |\theta_1(\frac{\sigma}{2\pi}|\frac{i\beta}{2\pi})|^2|\theta_2(\frac{\beta+4it}{4\pi i}|\frac{i\beta}{2\pi})|^2
    } \right)^{\frac{k^2}{N^2}}\,.   
\end{equation}
From Eq.~(\ref{twist2pt}) and using Eq.~(\ref{replica}), it's straightforward to obtain the entanglement entropy as 
\begin{equation}\label{SACFT}
    S_A(t, \sigma) = \frac{1}{6}\log\frac{
        |\theta_1(\frac{\sigma}{2\pi}|\frac{i\beta}{2\pi})|^2|\theta_2(\frac{\beta+4it}{4\pi i}|\frac{i\beta}{2\pi})|^2
    }{
        \eta(\frac{i\beta}{2\pi})^6 |\theta_2(\frac{\beta+4it}{4\pi i}+\frac{\sigma}{2\pi}|\frac{i\beta}{2\pi})||\theta_2(\frac{\beta+4it}{4\pi i}-\frac{\sigma}{2\pi}|\frac{i\beta}{2\pi})|
    }\,.
\end{equation}
\subsection{Proof of Eq.~(\ref{identity})}
Since the identity (\ref{identity}) is extremely crucial in our derivation of the twist operator correlator on the Klein bottle, here we provide its proof. We want to prove the following identity for bosonic creation and annihilation operators satisfying $[\hat{\alpha}, \hat{\alpha}^\dagger] = 1$ and $[\hat{\beta}, \hat{\beta}^\dagger] = 1$:
\begin{equation}
    I\equiv\langle 0| e^{\hat{\alpha}\hat{\beta}z} e^{a\hat{\alpha}+\bar{a}\hat{\beta}} e^{b\hat{\alpha}^\dagger+\bar{b}\hat{\beta}^\dagger} e^{\hat{\alpha}^\dagger\hat{\beta}^\dagger} |0\rangle = \frac{1}{1-z} \exp\left[\frac{ab+\bar{a}\bar{b}+a\bar{a}+zb\bar{b}}{1-z}\right]\,.
\end{equation}
We can use the two-mode \textit{coherent state} $\ket{\alpha,\beta}$ satisfying $\hat{\alpha}\ket{\alpha,\beta}=\alpha\ket{\alpha,\beta},\hat{\beta}\ket{\alpha,\beta}=\beta\ket{\alpha,\beta}$ to transform the left hand side of the identity above to an integral by inserting a completeness relation for two-mode coherent states, $1 = \int \frac{d^2\alpha \, d^2\beta}{\pi^2} |\alpha, \beta\rangle\langle\alpha, \beta|$, into the middle of the operator string. This converts the operator expectation value into a Gaussian integral over c-number variables $\alpha$ and $\beta$.

We insert the completeness relation to split the expression into two distinct matrix elements
\begin{equation}
    I = \int \frac{d^2\alpha \, d^2\beta}{\pi^2} \langle 0| e^{\hat{\alpha}\hat{\beta}z} e^{a\hat{\alpha} + \bar{a}\hat{\beta}} |\alpha, \beta\rangle \langle\alpha, \beta| e^{b\hat{\alpha}^\dagger + \bar{b}\hat{\beta}^\dagger} e^{\hat{\alpha}^\dagger\hat{\beta}^\dagger} |0\rangle\,.
\end{equation}
We now evaluate the two matrix elements in the integrand separately. The first one is
\begin{align}
    M_L = \langle 0| e^{\hat{\alpha}\hat{\beta}z + a\hat{\alpha} + \bar{a}\hat{\beta}} |\alpha, \beta\rangle= e^{z\alpha\beta + a\alpha + \bar{a}\beta} \langle 0 | \alpha, \beta \rangle\,.
\end{align}
Using the standard overlap formula of coherent state $\langle 0 | \alpha, \beta \rangle = e^{-\frac{1}{2}(|\alpha|^2 + |\beta|^2)}$, we get
\begin{equation}
    M_L = e^{z\alpha\beta + a\alpha + \bar{a}\beta} e^{-\frac{1}{2}(|\alpha|^2 + |\beta|^2)}\,.
\end{equation}
Similarly, for the second matrix element, we have
\begin{align}
    M_R = \langle \alpha, \beta | e^{b\hat{\alpha}^\dagger + \bar{b}\hat{\beta}^\dagger + \hat{\alpha}^\dagger\hat{\beta}^\dagger} |0\rangle= e^{b\alpha^* + \bar{b}\beta^* + \alpha^*\beta^*} \langle \alpha, \beta | 0 \rangle\,.
\end{align}
Using the overlap formula $\langle \alpha, \beta | 0 \rangle = e^{-\frac{1}{2}(|\alpha|^2 + |\beta|^2)}$, we obtain
\begin{equation}
    M_R = e^{b\alpha^* + \bar{b}\beta^* + \alpha^*\beta^*} e^{-\frac{1}{2}(|\alpha|^2 + |\beta|^2)}\,.
\end{equation}

\par We now substitute $M_L$ and $M_R$ back into the integral for $I$, obtaining
\begin{equation}
    I = \int \frac{d^2\alpha}{\pi} \frac{d^2\beta}{\pi} \exp\left[ -|\alpha|^2 - |\beta|^2 + z\alpha\beta + \alpha^*\beta^* + a\alpha + \bar{a}\beta + b\alpha^* + \bar{b}\beta^* \right]\,.
\end{equation}
We can solve this by iterated integration, first over $\alpha$ and then over $\beta$.

\paragraph{Integrating over $\alpha$}
To integrate over $\alpha$, we collect all terms in the exponent involving $\alpha, \alpha^*$ and complete the square as
\begin{equation}
    -\alpha^*\alpha + (a+z\beta)\alpha + (b+\beta^*)\alpha^* = -(\alpha - (b+\beta^*))(\alpha^* - (a+z\beta)) + (a+z\beta)(b+\beta^*)\,.
\end{equation}
Since $\int \frac{d^2\alpha}{\pi} e^{-(\alpha - J)(\alpha^* - K)} = 1$, integrating over $\alpha$ leaves the term $\exp\left[ (a+z\beta)(b+\beta^*) \right]$.

\paragraph{Integrating over $\beta$}
We combine the result from the $\alpha$-integration with the remaining $\beta$-dependent terms. The exponent for the $\beta$ integral is
\begin{align}
    (a+z\beta)(b+\beta^*) - |\beta|^2 + \bar{a}\beta + \bar{b}\beta^*= ab - (1-z)|\beta|^2 + (\bar{a}+zb)\beta + (a+\bar{b})\beta^*\,.
\end{align}
The remaining integral is
\begin{equation}
    I = e^{ab} \int \frac{d^2\beta}{\pi} \exp\left[ -(1-z)|\beta|^2 + (\bar{a}+zb)\beta + (a+\bar{b})\beta^* \right]\,.
\end{equation}
Using the Gaussian integral formula $\int \frac{d^2\zeta}{\pi} e^{-A|\zeta|^2 + J\zeta + K\zeta^*} = \frac{1}{A} e^{JK/A}$, with $A = 1-z$, $J = \bar{a}+zb$, and $K = a+\bar{b}$, we obtain
\begin{equation}
    I = e^{ab} \cdot \frac{1}{1-z} \exp\left[\frac{(\bar{a}+zb)(a+\bar{b})}{1-z}\right]=\frac{1}{1-z} \exp\left[\frac{ab + \bar{a}\bar{b} + a\bar{a} + zb\bar{b}}{1-z}\right]\,.
\end{equation}
This result perfectly matches the right-hand side of the identity we set out to prove. 
\section{Imaginary-time vs. real-time evolution of the fermionic covariance matrix}
In this section we give a complete derivation of the fermionic covariance matrix under imaginary-time and real-time evolution. 
\subsection{The imaginary-time evolution}
Consider an arbitrary Gaussian state $\rho(0)$ and evolve it according to
\begin{equation} 
\rho(\tau)=\frac{e^{-H\tau}\rho(0)e^{-H\tau}}{\text{Tr}(e^{-2H\tau}\rho(0))}\,.
\end{equation}
From the equation above, we can obtain the equation of motion for $\rho(\tau)$ for any Hamiltonian $H$ as
\begin{equation} 
\frac{d\rho}{d\tau}=-\{H,\rho(\tau)\}+2\rho(\tau)\text{Tr}(\rho(\tau)H)\,,
\end{equation}
where $\{A,B\}=AB+BA$ is the anti-commutator.

Then the evolution equation for the covariance matrix $\Gamma$ is
\begin{equation}\label{dGammadtau}
\frac{d\Gamma_{mn}(\tau)}{d\tau} = -\frac{i}{2}\text{Tr}\left( \rho(\tau)\{[\gamma_m, \gamma_n], H\} \right)+2\Gamma_{mn}(\tau)\text{Tr}( \rho(\tau)H)\,.
\end{equation}
We need to evaluate the expectation value of the operator $\{[\gamma_m, \gamma_n], H\}$. Let us expand it as
\begin{equation}\label{commute}
\{[\gamma_m, \gamma_n], H\} = (\gamma_m\gamma_n - \gamma_n\gamma_m)H + H(\gamma_m\gamma_n - \gamma_n\gamma_m)\,.
\end{equation}
We can calculate the expectation value of each of the four terms (e.g., $\langle \gamma_m\gamma_n H \rangle$) separately. Let us focus on the first term and substitute the Hamiltonian $H = \frac{i}{4} \sum_{k,l} \mathcal{H}_{kl} \gamma_k \gamma_l$, obtaining
\begin{equation}
\langle \gamma_m \gamma_n H \rangle = \frac{i}{4} \sum_{k,l} \mathcal{H}_{kl} \langle \gamma_m \gamma_n \gamma_k \gamma_l \rangle\,.
\end{equation}
This quantity is the expectation value of four Majorana operators. For any Gaussian state, it can be reduced to two-point correlators $M_{ab}\equiv\langle \gamma_a \gamma_b \rangle$ by Wick’s theorem.
The two-point functions are directly related to the covariance matrix as
\begin{equation}
\langle \gamma_a \gamma_b \rangle = -i \Gamma_{ab} + \delta_{ab}\,,
\end{equation}
or equivalently, $M=-i\Gamma + I$. Now, we apply Wick's theorem to $\langle \gamma_m \gamma_n \gamma_k \gamma_l \rangle$
\begin{equation}
\langle \gamma_m \gamma_n \gamma_k \gamma_l \rangle = \langle\gamma_m\gamma_n\rangle\langle\gamma_k\gamma_l\rangle - \langle\gamma_m\gamma_k\rangle\langle\gamma_n\gamma_l\rangle + \langle\gamma_m\gamma_l\rangle\langle\gamma_n\gamma_k\rangle\,.
\end{equation}
Let us substitute this into the expression for $\langle \gamma_m \gamma_n H \rangle$, we get
\begin{equation}
\frac{i}{4} \sum_{k,l} \mathcal{H}_{kl} \left( \langle\gamma_m\gamma_n\rangle\langle\gamma_k\gamma_l\rangle - \langle\gamma_m\gamma_k\rangle\langle\gamma_n\gamma_l\rangle + \langle\gamma_m\gamma_l\rangle\langle\gamma_n\gamma_k\rangle \right)\,.
\end{equation}
This looks complicated, but we can simplify by using the properties of the matrices. Since the matrix $\mathcal{H}$ is anti-symmetric.
The first term is proportional to $M_{mn}\text{Tr}(\mathcal{H}^T M) = -M_{mn}\text{Tr}(\mathcal{H}M)$. The second and third term are all proportional to $-(M\mathcal{H}M^{T})_{mn}$. Thus, we have
\begin{equation}\label{term1} 
\begin{split}
&\langle \gamma_m \gamma_n H \rangle=\frac{i}{4}(-M_{mn}\text{Tr}(\mathcal{H}M)-2(M\mathcal{H}M^{T})_{mn})\\
&=-\frac{i}{4}(M_{mn}\text{Tr}(\mathcal{H}M)+2(\Gamma\mathcal{H}\Gamma)_{mn}-2i(\Gamma\mathcal{H})_{mn}+2i(\mathcal{H}\Gamma)_{mn}+2\mathcal{H}_{mn})\,, 
\end{split}
\end{equation}
Similarly, we have
\begin{equation}\label{term2} 
\begin{split}
&\langle H\gamma_m \gamma_n  \rangle=\frac{i}{4}(-M_{mn}\text{Tr}(\mathcal{H}M)-2(M^{T}\mathcal{H}M)_{mn})\\
&=-\frac{i}{4}(M_{mn}\text{Tr}(\mathcal{H}M)+2(\Gamma\mathcal{H}\Gamma)_{mn}+2i(\Gamma\mathcal{H})_{mn}-2i(\mathcal{H}\Gamma)_{mn}+2\mathcal{H}_{mn})\,.  
\end{split}
\end{equation}
For $m\neq n$, $M_{mn}=-i\Gamma_{mn}, M_{mn}=-M_{nm}$, and  $-\frac{i}{4}\text{Tr}(\mathcal{H}M)=\text{Tr}(\rho(\tau)H)$. The other two terms from Eq.~(\ref{commute}) are obtained from Eq.~(\ref{term1}) and Eq.~(\ref{term2}) by exchanging $m$ and $n$.
Combining all the contributions from these four terms, we arrive at the result
\begin{equation}
\langle \{[\gamma_m, \gamma_n], H\} \rangle = -2i(\mathcal{H}_{mn} + (\Gamma\mathcal{H}\Gamma)_{mn})-4i\Gamma_{mn}\text{Tr}(\rho(\tau)H)\,.
\end{equation}
Then substitute the result above into Eq.~(\ref{dGammadtau}), we obtain the following imaginary time equation of motion \cite{kraus2010,Ashida:2018}
\begin{equation}\label{Riccati}
\frac{d\Gamma(\tau)}{d\tau} = -\mathcal{H}-\Gamma(\tau)\mathcal{H}\Gamma(\tau)\,.
\end{equation}
\par Let us make a simple consistent check of this equation. We should verify that our derived equation of motion (\ref{Riccati}) is consistent with the purity constraint. That is, if the condition $\Gamma^2=-I$ holds at $\tau=0$, it must hold for all $\tau > 0$. To prove this, we show that the time derivative of $\Gamma^2$ is zero if $\Gamma$ satisfies the equation of motion. We compute the derivative of $\Gamma^2$ using the chain rule
\begin{equation}
    \frac{d}{d\tau}(\Gamma^2) = \frac{d\Gamma}{d\tau}\Gamma + \Gamma\frac{d\Gamma}{d\tau}\,.
\end{equation}
Next, we substitute the expression for $\frac{d\Gamma}{d\tau}$ from the Riccati equation (\ref{Riccati})
\begin{equation}
\begin{split}
    &\frac{d}{d\tau}(\Gamma^2) = -\mathcal{H}\Gamma - \Gamma\mathcal{H}\Gamma^2 - \Gamma\mathcal{H} - \Gamma^2\mathcal{H}\Gamma\\
    &= -\mathcal{H}\Gamma - \Gamma\mathcal{H}(-I) - \Gamma\mathcal{H} - (-I)\mathcal{H}\Gamma=0\,,
\end{split}
\end{equation}
where we have used the purity condition, $\Gamma^2 = -I$, which holds at time $\tau=0$.
Since the time derivative of $\Gamma^2$ is identically zero, the property $\Gamma^2 = -I$ is a constant of motion. 
\subsection{The real-time evolution}
The real-time evolution of an arbitrary density matrix is given by
\begin{equation}
\rho(t)=e^{-iHt}\rho(0)e^{iHt},
\end{equation}
which is equivalent to
\begin{equation} 
\frac{d\rho(t)}{dt}=-i[H,\rho(t)]\,.
\end{equation}
We want to derive the formula that governs the time evolution of the fermionic covariance matrix $\Gamma(t)$ for a system evolving under a quadratic Hamiltonian $H$. The state of the system is $|\Psi(t)\rangle = e^{-iHt}|\Psi_0\rangle$, where $|\Psi_0\rangle$ is a pure Gaussian state (like the TPQ state) with a known initial covariance matrix $\Gamma(0)$.

The most direct and unambiguous way to derive the evolution of an expectation value is through the Heisenberg picture, where the quantum state remains fixed at $|\Psi_0\rangle$, and the operators evolve in time.
The evolution of any operator $\mathcal{O}(t)$ in the Heisenberg picture is given by
\begin{equation}
\mathcal{O}(t) = e^{iHt} \mathcal{O}(0) e^{-iHt}\,.
\end{equation}
Its equation of motion is the Heisenberg equation (setting $\hbar=1$)
\begin{equation}
i \frac{d\mathcal{O}(t)}{dt} = [\mathcal{O}(t), H]\,.
\end{equation}
We apply this to the vector of Majorana fermion operators $\vec{\gamma}(t)$
\begin{equation}
i \frac{d\vec{\gamma}(t)}{dt} = [\vec{\gamma}(t), H]\,.
\end{equation}
This is the bridge that connects the many-body Hamiltonian $H$ to the single-particle matrix $\mathcal{H}$. We need to compute the commutator $[H, \gamma_m]$.
The Hamiltonian is $H = \frac{i}{4} \sum_{k,l} \mathcal{H}_{kl} \gamma_k \gamma_l$.
\begin{equation}
[H, \gamma_m] = \left[ \frac{i}{4} \sum_{k,l} \mathcal{H}_{kl} \gamma_k \gamma_l, \gamma_m \right] = \frac{i}{4} \sum_{k,l} \mathcal{H}_{kl} [\gamma_k \gamma_l, \gamma_m]\,.
\end{equation}
Using the identity $[\text{AB, C}] = \text{A}\{\text{B,C}\} - \{\text{A,C}\}\text{B}$ and the Majorana anti-commutator $\{\gamma_a, \gamma_b\} = 2\delta_{ab}$, we get
\begin{equation}
[\gamma_k \gamma_l, \gamma_m] = 2\delta_{lm}\gamma_k - 2\delta_{km}\gamma_l\,.
\end{equation}
Substituting this back into the sum, we obtain
\begin{equation}
[H, \gamma_m] = \frac{i}{4} \sum_{k,l} \mathcal{H}_{kl} (2\delta_{lm}\gamma_k - 2\delta_{km}\gamma_l) = \frac{i}{2} \left( \sum_{k} \mathcal{H}_{km}\gamma_k - \sum_{l} \mathcal{H}_{ml}\gamma_l \right)\,.
\end{equation}
Using the anti-symmetry of the Majorana Hamiltonian matrix, $\mathcal{H}_{km} = -\mathcal{H}_{mk}$, we have
\begin{equation}
[H, \gamma_m] = \frac{i}{2} \left( -\sum_{k} \mathcal{H}_{mk}\gamma_k - \sum_{l} \mathcal{H}_{ml}\gamma_l \right) = -i \sum_k \mathcal{H}_{mk} \gamma_k\,.
\end{equation}
In matrix notation, the commutator is simply
\begin{equation}
[H, \vec{\gamma}] = -i\mathcal{H}\vec{\gamma}\,.
\end{equation}
Now substituting this commutator back into the Heisenberg equation of motion, we have
\begin{equation}
i \frac{d\vec{\gamma}(t)}{dt} = [\vec{\gamma}(t), H]\,.
\end{equation}
Since the commutator has the same form for the time-evolved operators, we have
\begin{equation}
i\frac{d\vec{\gamma}(t)}{dt} = i \mathcal{H} \vec{\gamma}(t)\,.
\end{equation}
The solution is given by the matrix exponential as
\begin{equation}
\vec{\gamma}(t) = e^{\mathcal{H}t} \vec{\gamma}(0)\,.
\end{equation}
Finally, we compute the covariance matrix at time $t$. Its definition involves the expectation value of the time-evolved operators in the initial state $|\Psi_0\rangle$.
\begin{equation}
\Gamma_{mn}(t) = \frac i 2\langle \Psi_0 | [\gamma_m(t), \gamma_n(t)] | \Psi_0 \rangle\,.
\end{equation}
In matrix form, the equation above reads
\begin{equation}
\Gamma(t) = \frac i 2 \langle [\vec{\gamma}(t), \vec{\gamma}(t)^T] \rangle_0\,.
\end{equation}
Now, substitute the solution of $\vec{\gamma}(t)$ to obtain
\begin{equation}
\Gamma(t) = \frac i 2 \langle [e^{\mathcal{H}t}\vec{\gamma}(0), (e^{\mathcal{H}t}\vec{\gamma}(0))^T] \rangle_0\,.
\end{equation}
The matrix exponential $e^{\mathcal{H}t}$ is a matrix of constant coefficients, so we can pull it out of the expectation value and the commutator
\begin{equation}
\Gamma(t) = e^{\mathcal{H}t} \left( \frac i 2 \langle [\vec{\gamma}(0), \vec{\gamma}(0)^T] \rangle_0 \right)(e^{\mathcal{H}t})^T=e^{\mathcal{H}t} \Gamma(0) (e^{\mathcal{H}t})^T\,.
\end{equation}
Since $\mathcal{H}$ is a real, anti-symmetric matrix, the matrix exponential $U(t) = e^{\mathcal{H}t}$ is a real, orthogonal matrix. An orthogonal matrix satisfies the property that its transpose is its inverse: $U^T = U^{-1}$.
\begin{equation}
(e^{\mathcal{H}t})^T =e^{-\mathcal{H}t}\,.
\end{equation}
Substituting this back into our expression for $\Gamma(t)$, we finally have
\begin{equation}
\Gamma(t) = e^{\mathcal{H}t} \Gamma(0) e^{-\mathcal{H}t}\,.
\end{equation}
\section{The solution of general matrix Riccati equations}
\par In this section, we outline the standard method for solving the most general matrix Riccati equation \cite{AbouKandil:2003,Bellman:1978}.
A standard form of the matrix Riccati differential equation for an $n \times n$ matrix $X(t)$ is
\begin{equation}\label{GRiccati}
\frac{dX(t)}{dt} = A + B X(t) + X(t) C + X(t) D X(t)\,,
\end{equation}
where $A, B, C, D$ are all $n \times n$ constant matrices. 
\par Our imaginary time evolution equation for the covariance matrix reads
\begin{equation}\label{riccati}
\frac{d\Gamma(\tau)}{d\tau} = -\mathcal{H} - \Gamma(\tau)\mathcal{H}\Gamma(\tau)\,,
\end{equation}
which can be obtained from the general Riccati equation (\ref{GRiccati}) by setting $A = -\mathcal{H}$, $B=C=\mathbf{0}$ (the zero matrix), and $D=-\mathcal{H}$.
The strategy is to find a solution of the form $X(t) = P(t)Q(t)^{-1}$, where the block matrix $\begin{pmatrix} P(t) \\ Q(t) \end{pmatrix}$ evolves according to a simple linear differential equation.
\par We propose that the solution to the Riccati equation can be written as the ratio of two matrices
\begin{equation}
X(t) = P(t) Q(t)^{-1}\,,
\end{equation}
where $P(t)$ is $n \times n$ and $Q(t)$ is an invertible $n \times n$ matrix. We now differentiate this expression using the product rule and the rule for the derivative of a matrix inverse ($\frac{d}{dt}(A^{-1}) = -A^{-1} \frac{dA}{dt} A^{-1}$):
\begin{equation}
\frac{dX}{dt} = \frac{d}{dt}(PQ^{-1}) = P'Q^{-1} + P(Q^{-1})' = P'Q^{-1} - PQ^{-1}Q'Q^{-1}\,.
\end{equation}
Now, substituting the ansatz $X = PQ^{-1}$ and its derivative into the original Riccati equation, we obtain
\begin{equation}
P'Q^{-1} - PQ^{-1}Q'Q^{-1} = A + B(PQ^{-1}) + (PQ^{-1})C + (PQ^{-1})D(PQ^{-1})\,.
\end{equation}
Multiplying the entire equation from the right by $Q(t)$, we get
\begin{equation}
P' - PQ^{-1}Q' = AQ + BP + PQ^{-1}CQ + PQ^{-1}DP\,.
\end{equation}

Let us group terms for $P'$ and $Q'$ and rearrange the equation as
\begin{equation}
P' = (BP + AQ) + PQ^{-1}(Q' + CQ + DP)\,.
\end{equation}
This equation is satisfied if we require the block vector $\begin{pmatrix} P(t) \\ Q(t) \end{pmatrix}$ evolves as 
\begin{equation}\label{linear}
\frac{d}{dt} \begin{pmatrix} P(t) \\ Q(t) \end{pmatrix} = \begin{pmatrix} B & A \\ -D & -C \end{pmatrix} \begin{pmatrix} P(t) \\ Q(t) \end{pmatrix}\,.
\end{equation}
This is now a linear system of differential equations. Let us write it out explicitly 
\begin{equation}
P' = BP + AQ, \qquad Q' = -DP - CQ\,.
\end{equation}
The solution to the linear system is given by the matrix exponential
\begin{equation}
\begin{pmatrix} P(t) \\ Q(t) \end{pmatrix} = \exp\left( t \begin{pmatrix} B & A \\ -D & -C \end{pmatrix} \right) \begin{pmatrix} P(0) \\ Q(0) \end{pmatrix}.
\end{equation}

Let us call the $2n \times 2n$ matrix $M = \begin{pmatrix} B & A \\ -D & -C \end{pmatrix}$, then 
\begin{equation}
\begin{pmatrix} P(t) \\ Q(t) \end{pmatrix} = e^{tM} \begin{pmatrix} P(0) \\ Q(0) \end{pmatrix}\,.
\end{equation}
Let us partition the matrix exponential $e^{tM}$ into four $n \times n$ blocks
\begin{equation}
e^{tM} = \begin{pmatrix} E_{11}(t) & E_{12}(t) \\ E_{21}(t) & E_{22}(t) \end{pmatrix}.
\end{equation}

The initial condition of the original equation (\ref{GRiccati}) is $X(0) = X_0$. We choose the initial conditions for the linear system (\ref{linear}) to satisfy this, for instance, by setting $P(0) = X_0$  and $Q(0) = I$ (the identity matrix). Then, the solution of $P$ and $Q$ is
\begin{equation} 
\begin{split}
&P(t) = E_{11}(t)P(0) + E_{12}(t)Q(0) = E_{11}(t)X_0 + E_{12}(t)\,, \\
&Q(t) = E_{21}(t)P(0) + E_{22}(t)Q(0) = E_{21}(t)X_0 + E_{22}(t)\,.
\end{split}
\end{equation}
The final solution to the original Riccati equation is given by
\begin{equation}
X(t) = \left( E_{11}(t)X_0 + E_{12}(t)\right) \left(E_{21}(t)X_0 + E_{22}(t)\right)^{-1}\,.
\end{equation}

For our problem, the linear system corresponding to Eq.~(\ref{riccati}) is
\begin{equation}
\frac{d}{d\tau} \begin{pmatrix} P(\tau) \\ Q(\tau) \end{pmatrix} = \begin{pmatrix} 0 & -\mathcal{H} \\ \mathcal{H} & 0 \end{pmatrix} \begin{pmatrix} P(\tau) \\ Q(\tau) \end{pmatrix}\,.
\end{equation}
The solution to this linear system is given by the matrix exponential
\begin{equation}
\begin{pmatrix} P(\tau) \\ Q(\tau) \end{pmatrix} = \exp\left( \tau \begin{pmatrix} 0 & -\mathcal{H} \\ \mathcal{H} & 0 \end{pmatrix} \right) \begin{pmatrix} P(0) \\ Q(0) \end{pmatrix}\,.
\end{equation}
The exponential of this block-matrix can be calculated easily. Let $M = \begin{pmatrix} 0 & -\mathcal{H} \\ \mathcal{H} & 0 \end{pmatrix}$, we have 
\begin{equation}
e^{\tau M} = \begin{pmatrix} \cos(\mathcal{H}\tau) & -\sin(\mathcal{H}\tau) \\ \sin(\mathcal{H}\tau) & \cos(\mathcal{H}\tau) \end{pmatrix}.
\end{equation}
So the evolved matrices are
\begin{equation}
\begin{split}
&P(\tau) = \cos(\mathcal{H}\tau)P(0) - \sin(\mathcal{H}\tau)Q(0)\,, \\
&Q(\tau) = \sin(\mathcal{H}\tau)P(0) + \cos(\mathcal{H}\tau)Q(0)\,.
\end{split}
\end{equation}
Now we specify the initial conditions by imposing $\Gamma_0 = P(0)Q(0)^{-1}$. Choosing $P(0)=\Gamma_0$ and $Q(0)=I$, we then obtain
\begin{equation}
\begin{split}
&P(\tau) = \cos(\mathcal{H}\tau)\Gamma_0 - \sin(\mathcal{H}\tau)\,, \\
&Q(\tau) = \sin(\mathcal{H}\tau)\Gamma_0 + \cos(\mathcal{H}\tau)\,.
\end{split}
\end{equation}
The final solution is obtained from $\Gamma(\tau) = P(\tau)Q(\tau)^{-1}$ as
\begin{equation}
\Gamma(\tau) = \left( \cos(\mathcal{H}\tau)\Gamma_0 - \sin(\mathcal{H}\tau) \right) \left(\sin(\mathcal{H}\tau)\Gamma_0 + \cos(\mathcal{H}\tau) \right)^{-1}\,.
\end{equation}

\section{Exact BCS Representation of the Crosscap State}

In this section, we provide a derivation of the momentum-space BCS representation of the crosscap state. By exploiting the translational symmetry inherent in the anti-periodic boundary conditions (APBC), we show that the state is strictly momentum-diagonal for any finite system size $N$.

Consider a free fermion chain of length $L=2N$ with APBC, $c_{j+2N} = -c_j$. The unnormalized crosscap state is defined in real space as $|\tilde{\mathcal{C}}\rangle = \exp(\hat{Q}_N)|0\rangle$, where the generator pairs antipodal sites:
\begin{equation}
\hat{Q}_N = \sum_{j=1}^N c_j^\dagger c_{j+N}^\dagger\,.
\end{equation}
To diagonalize this generator in momentum space, we introduce an auxiliary operator summing over the entire lattice:
\begin{equation}
\hat{Q}_{2N} = \sum_{j=1}^{2N} c_j^\dagger c_{j+N}^\dagger = \sum_{j=1}^N c_j^\dagger c_{j+N}^\dagger + \sum_{j=N+1}^{2N} c_j^\dagger c_{j+N}^\dagger\,.
\end{equation}
By shifting the summation index $l = j - N$ in the second term and applying the APBC ($c_{l+2N}^\dagger = -c_l^\dagger$), we obtain
\begin{equation}
\sum_{l=1}^N c_{l+N}^\dagger c_{l+2N}^\dagger = -\sum_{l=1}^N c_{l+N}^\dagger c_l^\dagger = \sum_{l=1}^N c_l^\dagger c_{l+N}^\dagger = \hat{Q}_N\,,
\end{equation}
where the last step utilizes the fermionic anticommutation relation $c_a^\dagger c_b^\dagger = -c_b^\dagger c_a^\dagger$. This yields  $\hat{Q}_{2N} = 2\hat{Q}_N$ or equivalently
\begin{equation}
\hat{Q}_N = \frac{1}{2} \hat{Q}_{2N} = \frac{1}{2} \sum_{j=1}^{2N} c_j^\dagger c_{j+N}^\dagger\,.
\end{equation}

We now substitute the Fourier transform $c_j^\dagger = \frac{1}{\sqrt{2N}} \sum_k e^{-ikj} c_k^\dagger$ into the full-lattice sum. The orthogonality of the Fourier modes over the $2N$ sites rigorously enforces momentum conservation:
\begin{equation}
\begin{aligned}
\hat{Q}_N &= \frac{1}{4N} \sum_{k,q} e^{-iqN} c_k^\dagger c_q^\dagger \sum_{j=1}^{2N} e^{-i(k+q)j} \\
&= \frac{1}{4N} \sum_{k,q} e^{-iqN} c_k^\dagger c_q^\dagger \left( 2N \delta_{k,-q} \right) \\
&= \frac{1}{2} \sum_k e^{ikN} c_k^\dagger c_{-k}^\dagger\,.
\end{aligned}
\end{equation}
Under APBC, the quantized momenta are $k = \frac{\pi}{N}(m + 1/2)$ with integer $m \in[-N, N-1]$. Thus, the phase factor simplifies exactly to $e^{ikN} = e^{i\pi m}e^{i\pi/2} = i(-1)^m$. Because the operator is strictly momentum-diagonal for any finite $N$, the normalized initial state is an exact BCS state:
\begin{equation}
|\mathcal{C}\rangle = \prod_{k>0} \frac{1}{\sqrt{2}} \left[ 1 + i(-1)^m c_k^\dagger c_{-k}^\dagger \right] |0\rangle\,.
\end{equation}

Applying the imaginary-time evolution $e^{-\beta H/4}$ to this state immediately yields the thermal pure quantum (TPQ) state $|\Psi_\beta\rangle$ used in the main text:
\begin{equation}
|\Psi_\beta\rangle = \prod_{k>0}\left[u_k(\beta)+v_k(\beta)c_k^\dagger c_{-k}^\dagger\right]|0\rangle\,,
\end{equation}
with the renormalized, $\beta$-dependent Bogoliubov coefficients:
\begin{equation}
|u_k(\beta)|^2 = \frac{1}{1 + e^{-\beta E(k)}}\,,\quad |v_k(\beta)|^2 = \frac{e^{-\beta E(k)}}{1 + e^{-\beta E(k)}}\,,
\end{equation}
and their ratio fixed by $v_k(\beta)/u_k(\beta) = i(-1)^m e^{-\beta E(k)/2}$. This exact finite-size representation provides the rigorous foundation for the independent quasiparticle picture governing the entanglement dynamics.
We finally get
\begin{equation}\label{FermiDirac}
n(k,\beta) = |v_k(\beta)|^2=\frac{1}{1+e^{\beta E(k)}}\,. 
\end{equation}
This formula correctly reproduces the Fermi-Dirac distribution at inverse temperature $\beta$.

\end{document}